\documentclass[aps,prb,showpacs,twocolumn,superscriptaddress]{revtex4}
\usepackage{amsmath,graphics,amssymb}
\usepackage[next]{inputenc}
\usepackage[dvips]{epsfig}
\def\be{\begin{equation}}
\def\ee{\end{equation}}

\def\bi{\begin{itemize}}
\def\ei{\end{itemize}}
\def\bn{\begin{enumerate}}
\def\en{\end{enumerate}}
\def\bea{\begin{eqnarray}}
\def\eea{\end{eqnarray}}
\def\no{\nonumber}
\def\ba{\begin{array}}
\def\ea{\end{array}}
\def\bd{\begin{displaymath}}
\def\ed{\end{displaymath}}

\begin{document}
\title{Dzyaloshinskii-Moriya Interaction and Anisotropy effects on the Entanglement of Heisenberg Model }

\author{M. Kargarian}
\affiliation{Physics Department, Sharif University of Technology,
Tehran 11155-9161, Iran}
\author{R. Jafari}
\affiliation{Institute for Advanced Studies in Basic Sciences,
Zanjan 45195-1159, Iran}
\author{A. Langari}
\affiliation{Physics Department, Sharif University of Technology,
Tehran 11155-9161, Iran} \email[]{langari@sharif.edu}

\begin{abstract}
In this paper the the effect of Dzyaloshinskii-Moriya interaction
and anisotropy on the Entanglement of Heisenberg model has been
studied. While the anisotropy suppress the entanglement due to
favoring of the alignment of spins, the DM interaction restore the
spoiled entanglement via the introduction of the quantum
fluctuations. Thermodynamic limit of the model and emerging of
non-analytic behavior of the entanglement have also been probed. The
singularities of the entanglement correspond to the critical
boundary separating different phases of the model. The effect of
gapped and gapless phases of the model on the features of the
entanglement has also been discussed.

\end{abstract}
\date{\today}


\maketitle
\section{Introduction \label{introduction}}
In recent years, the notion of entanglement has received much
attention in quantum information theory due to its notable features
in developing the idea of quantum computers and other quantum
information devices. Entanglement is a purely quantum correlation
without classical counterpart\cite{Bell} and has been realized as a
crucial resource to process and send information in novel ways such
as quantum teleportation, supercoding and algorithms for quantum
computations\cite{Nielsen}. Concerning the correlation content of
the entanglement, states of systems in condensed matter physics may
deserve the investigation of entanglement as a unique measure of
quantum correlations. The interest will be intensified when we
consider the relation between entanglement and quantum phase
transition where a drastic change in the ground state of the system
occur\cite{Sachdev}. This change will occur at zero temperature
where all thermal fluctuations get frozen and only surviving quantum
fluctuations drive the phase transition. In the past few years the
subject of many activities  were to investigate the role of
entanglement in the vicinity of quantum critical point for different
spin models \cite{Osterloh,Wu,Vidal1,Vidal2,Bose,Osborne,Verstraete,
Amico,Gu,Anfossi}.

Spin models provide not only a test ground for above issue but also
as a play-ground for implementation of may quantum information
protocols\cite{loss,raussendorf}. Among them are Ising model in
transverse field (ITF) and XXZ models. Despite of their simple
Hamiltonian, low energy behavior of many systems can be captured
through them. Ising model in transverse field has the benefit of
exact solvability through mapping to the free fermion
model\cite{Sachdev}. Such solvability provide the possibility to
test the behavior of entanglement and its scaling close to the
critical point of the system and perform a finite size scaling as
has been done in the seminal work of Osterloh and his
coworkers\cite{Osterloh}. The scenario is different in the XXZ model
where the entanglement between the two nearest-neighbor sites
develops a maximum at the isotropic point ($\Delta=1$) without any
singularity in its first derivative \cite{Shi3} which vanishes at
the critical point $\Delta=1$. However, the block-block entanglement
\cite{lou} of the spin-1/2 XX model with three-spin and uniform long
range interaction shows a logarithmic and algebraic dependence on
the size of block for different phases. Logarithmic divergences of
the entanglement entropy is a general feature of all one-dimension
critical systems where the coefficient of the logarithm is just the
central charge of the underlying critical theory\cite{cardy}.

The scaling of entanglement close to the phase transition and its
connection to the universality class of the model can be further
investigated through employing the renormalization group. This
method as we will see in the next sections provides a rather
analytic framework for treating the phases of the model even for
those that are beyond the exact solution. In this stream the scaling
of the entanglement govern by the critical exponent of the
model\cite{kargarian1, kargarian2}. However, The renormalization of
quantum states has also been introduced in terms of matrix product
states \cite{fv}.

Both Ising and XXZ models can be supplemented with a magnetic term,
the so called Dzyaloshinskii-Moriya (DM) interaction,  arising from
the spin-orbit coupling. Based on the symmetry
aspects\cite{Dzyaloshinskii}, It can be derived microscopically as a
linear correction to the standard superexchange
mechanism\cite{Moriya}. The interaction has the form
$\sum_{<ij>}\overrightarrow{D}_{ij}.(\overrightarrow{S_{i}}\times\overrightarrow{S_{j}})$
where the sum is over the pairs of spins. Some quantum
antiferromagnetic (AF) systems are expected to be described by DM
interaction with the underlying helical magnetic structures. Ising
model with DM interaction was extensively studied\cite{jafari1}. DM
interaction drive the quantum fluctuations resulting in a phase
transition in the model. Critical point separates the
antiferromagnetic and chiral phases. Derivative of the entanglement
diverge at the critical point with the critical exponent of the
model.

In this paper we address the behavior of the entanglement in the XXZ
model with DM interaction. Including the DM interaction makes the
phase diagram rich with a critical line instead a single
point\cite{jafari2}. First we employ the quantum renormalization
group to have a tractable problem. Afterwards, the entanglement
between degrees of freedom is treated through the renormalization.
We will see that derivative of the entanglement becomes singular at
the phase boundary and its scaling correspond to the gapless and
gapped phases of the model. The organization of the paper is as
follows. In the next section we briefly introduce the
renormalization group approach. In the Sec(\ref{entanglement}) we
exemplifies the effect of anisotropy and DM interaction. Then in
Sec(\ref{scaling}) we turn on to discuss the scaling of the
entanglement, and the Sec(\ref{conclusion}) is devoted to the
conclusions.

\section{Quantum renormalization group \label{qrg}}

The Quantum renormalization group presents a tractable version of
treating quantum systems at zero temperatures. Through the
renormalization the original model Hamiltonian is replaced by an
effective one in cost of renormalizing coupling constants. In this
way the original Hilbert space truncated to an reduced Hilbert space
including the effective degrees of freedom. Getting ride of extra
degrees of freedom gives rises to the flow of the coupling constants
in the parameter space of the model. The version we employ to kill
the degrees of freedom is Kadanoff's block approach since it is well
suited to perform analytical calculations in the lattice models and
they are conceptually easy to be extended to the higher
dimensions\cite{miguel1,miguel-book,Langari,jafari3}. In the
Kadanoff's method, the lattice is divided into blocks in which the
Hamiltonian is exactly diagonalized. By selecting a number of
low-lying eigenstates of the blocks the full Hamiltonian is
projected onto these eigenstates giving the effective (renormalized)
Hamiltonian.

The Hamiltonian of XXZ model with DM interaction in the $z$
direction on a periodic chain of $N$ sites is \bea \no
H(J,\Delta)=\frac{J}{4}\sum_{i}^{N}\Big[\sigma_{i}^{x}\sigma_{i+1}^{x}+\sigma_{i}^{y}\sigma_{i+1}^{y}+
\Delta\sigma_{i}^{z}\sigma_{i+1}^{z}\\
+D(\sigma_{i}^{x}\sigma_{i+1}^{y}-\sigma_{i}^{y}\sigma_{i+1}^{x})\Big]~~~D,J,\Delta>0
\eea

We divide the chain into blocks each containing three sites. The
block Hamiltonian $(H_{B}=\sum h_{I}^{B})$ of the three sites and
its eigenstates and eigenvalues are given in Appendix A of
reference\cite{jafari2}. However, we give only the degenerate ground
states since we need for them for evaluation of entanglement and
subsequent discussions, as follows:
\bea \nonumber |\psi_{0}\rangle
=\frac{1}{\sqrt{2q(q+\Delta)(1+D^{2})}}\Big(2(D^{2}+1)|\downarrow\downarrow\uparrow\rangle\\-
(1-iD)(\Delta+q)|\downarrow\uparrow\downarrow\rangle-2[2iD+(D^{2}-1)]|\uparrow\downarrow\downarrow\rangle\Big),\\
\no
|\psi_{0}'\rangle=\frac{1}{\sqrt{2q(q+\Delta)(1+D^{2})}}\Big(2(D^{2}+1)|\downarrow\uparrow\uparrow\rangle\\-
(1-iD)(\Delta+q)|\uparrow\downarrow\uparrow\rangle-2[2iD+(D^{2}-1)]|\uparrow\uparrow\downarrow\rangle)\Big),
\eea
where $|\uparrow\rangle$ and $|\downarrow\rangle$ are eigenstates of
the $\sigma^{z}$ Pauli operator and
$q=\sqrt{\Delta^{2}+8(1+D^{2})}$. The projection operator of the
ground state subspace defined by
$\big(P_{0}=|\Uparrow\rangle\langle\psi_{0}|+|\Downarrow\rangle\langle\psi_{0}'|\big)$,
Where $|\psi_{0}\rangle$ and $|\psi_{0}'\rangle$ are the doubly
degenerate ground states, $|\Uparrow\rangle$ and
$|\Downarrow\rangle$ are the renamed base kets in the effective
Hilbert space. We have kept two states ($|\psi_{0}\rangle$ and
$|\psi_{0}'\rangle$) for each block to define the effective (new)
site. Thus, the effective site can be considered as a spin
$\frac{1}{2}$. The effective Hamiltonian is similar to the initial
one, i.e, \bea \no
H^{eff}&=&\frac{J'}{4}\sum_{i}^{N}\Big[\sigma_{i}^{x}\sigma_{i+1}^{x}+\sigma_{i}^{y}\sigma_{i+1}^{y}+
{\Delta'}\sigma_{i}^{z}\sigma_{i+1}^{z}\\
&&+D'(\sigma_{i}^{x}\sigma_{i+1}^{y}-\sigma_{i}^{y}\sigma_{i+1}^{x})\Big]
\eea where $J'$ and $D'$ are the renormalized coupling constants.
The renormalized coupling constants are functions of the original
ones which are given by the following equations. \bea \label{rg}
J'=J(\frac{2}{q})^{2}(1+D^{2}),~~\Delta'=\frac{\Delta}{1+D^{2}}(\frac{\Delta+q}{4})^{2},~~D'=D.
\eea The above RG equations show that there is a phase boundary
$\Delta_{c}=\sqrt{1+D^{2}}$  that separates the spin fluid phase,
$\Delta<\sqrt{1+D^{2}}$, from the N\'{e}el phase,
$\Delta>\sqrt{1+D^{2}}$~~\cite{jafari2}.
\begin{figure}
\begin{center}
\includegraphics[width=8cm]{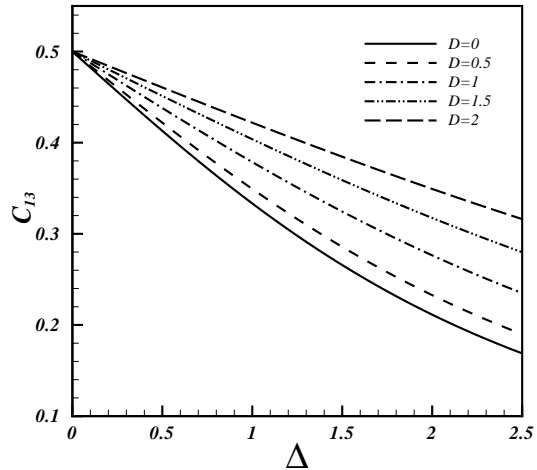}
\caption{Concurrence between first and third sites of a three-site
model in terms of anisotropy for different values of DM
interactions.} \label{fig1}
\end{center}
\end{figure}

\section{Entanglement analysis  \label{entanglement}}
Many measures of entanglement have been introduced and
analyzed\cite{Hill,Wootters,Vedral,Horodecki}, but the most relevant
to this work is the "entanglement of formation". For a reduced
density matrix $\rho_{ij}$ of two qubits that arises after
integrating out other degrees of freedom, the entanglement between
two qubits is evaluated as
$E=h(\frac{1}{2}+\frac{1}{2}\sqrt{1-C^{2}})$, where $h$ is a binary
entropy function $h(x)=-x\log_{2}x-(1-x)\log_{2}(1-x)$ and $C$
denotes the concurrence\cite{Wootters} defined as \bea \label{eqcon}
C=Max\{\lambda_{1}-\lambda_{2}-\lambda_{3}-\lambda_{4},0\},\eea
where $\lambda_{k}(k=1,2,3,4)$ are the square roots of the
eigenvalues in descending order of the operator $R_{ij}$: \bea \no
R_{ij}=\rho_{ij}\tilde{\rho}_{ij}~~~,~~~ \tilde{\rho}_{ij}=
(\sigma^{y}_{1}\otimes\sigma^{y}_{2})\rho_{ij}^{\ast}(\sigma^{y}_{1}\otimes\sigma^{y}_{2}).
\eea In this section we consider only a three site block and study
the effect of DM interaction and anisotropy parameter, i.e $D$ and
$\Delta$, respectively, on the entanglement between two spins
located on the sides of the block. To this end, let
$|\psi_{0}\rangle$ be the ground state of the block. By tracing the
density matrix $\rho=|\psi_{0}\rangle|\psi_{0}\rangle$ with respect
to the middle site of the block, the obtained reduced density matrix
and Eq.(\ref{eqcon}) give the expression for the concurrence in
terms of couplings $D$ and $\Delta$.

For different values of DM interaction and anisotropy parameters,
the plots of concurrence between first and third sites of the block
$C_{13}$ have been depicted in Fig.(\ref{fig1}). Consider first the
case of $D=0$. In this case the model becomes the known $XXZ$ model.
Large value of $\Delta$ implies the Neel state. Naturally this state
is an product state without any entanglement between its
constituents. As the anisotropy parameter reduces the quantum
fluctuations arising from the transverse interactions have dominant
effect and destroy Neel state. Indeed, the in-planar interactions
drive the quantum correlations, i.e the qubits in the presence of
quantum fluctuations are quantum correlated. The main message of the
Fig.(\ref{fig1}) is that the suppression of the entanglement can be
compensated by tuning the DM interaction. In the absence of the
anisotropy, the entanglement is insensitive to the DM value since
both transverse interaction and DM term stimulate the quantum
fluctuations. For nonzero value of anisotropy is clearly seen that
the turning on the DM interaction restores the spoiled entanglement.
The emerging of Neel state at large value of anisotropy and dominant
quantum fluctuations at small value tempt to conclude that in the
thermodynamic limit of the model there may occur quantum phase
transition with the critical boundary depend on the competition
between the parameters in the Hamiltonian. We will address this
issue in the next section.

\section{thermodynamic limit and non-analytic behavior of entanglement \label{scaling}}
In this section we would like to see that how the quantum phase
transition in the model, which can be signaled as unstable fixed
point of RG equations, can be realized by examining the behavior of
the entanglement. Indeed, the non-analytic behavior in some physical
quantity is a feature of second-order quantum phase transition. It
is also accompanied by a scaling behavior since the correlation
length diverges and there is no characteristic length scale in the
system at the critical point. As we already pointed out the
renormalization group allows us to capture the thermodynamic
properties of the model by considering a block of a few sites that
is analytically tractable. In fact the global properties of the
model enter a few sites through the renormalizing of coupling
constants. We exploit this advantage to study the scaling of the
entanglement in the model. Notice that in the $n-th$ step of RG a
system with size $n^{n+1}_{B}$ ($n_{B}$ is the number of sites in
each block) describes effectively by a model consisting of only
$n_{B}$ sites with the renormalized of coupling constants. The case
of the XXZ model has been extensively studied\cite{kargarian2},
where the critical point $\Delta=1$ separates spin-fluid and Neel
phases. However, for the present model the contribution of the
planar DM interaction tune the critical point of the model due to
involving the quantum fluctuations. Since the DM interaction doesn't
flow, as is clear from the RG equations in Eq.(\ref{rg}), it can be
treated as a fixed parameter. Now we put the next step forward to
see the evolution of the entanglement as the size of the system
becomes large through the RG steps.

\begin{figure}
\begin{center}
\includegraphics[width=8cm]{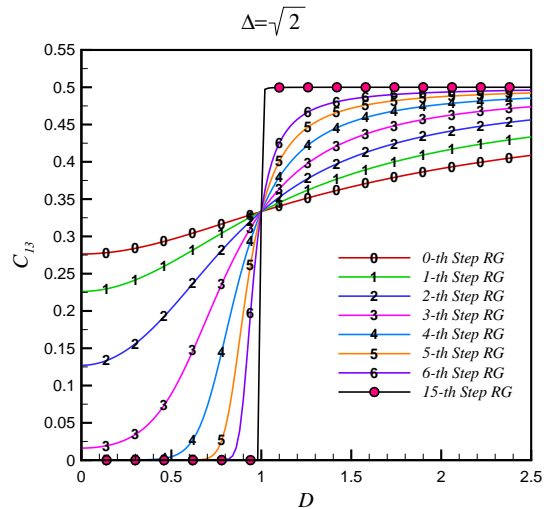}
\caption{(color online) Representation of the evolution of
entanglement entropy in terms of RG iterations at fixed value of
anisotropy $\Delta=\sqrt{2}$. Two different behavior of the
entanglement at the large steps of RG correspond to the emerging
phases of the model through the phase transition.} \label{fig2}
\end{center}
\end{figure}
Zero-step RG represents a three-site model that its entanglement
studied in the preceding section. However, the first-step RG stands
for a nine-site model which effectively describes by a three-site
model in the cost of renormalized coupling constants. In that case
the entanglement measures the correlation between effective degrees
of freedom. In each step of RG we can see the variation of the
entanglement in terms of anisotropy parameter with the fixed value
of DM interaction. All these data have been shown in
Fig.(\ref{fig2}). In this figure we have set the $\Delta=\sqrt{2}$.
It reveals that as the thermodynamic limit is touched via the
increasing of the RG steps, the entanglement develops two rather
different features. Indeed, there is a value for $D=1$ that
separates the different features. This value is exactly the critical
point of the model which is consistent with
$\Delta_{c}=\sqrt{1+D^{2}}$ if we set $\Delta=\sqrt{2}$. Different
features of the entanglement correspond to the emerging phases on
both sides of the critical point. For anisotropy parameter larger
than the critical point the Neel ordering dominates the phase of the
model while for anisotropy parameter less than the critical value
the increasing of the planar quantum fluctuations spoil any magnetic
ordering. This feature is not a specific character of the model
arising at $\Delta=\sqrt{2}$. In fact for any value anisotropy
$\Delta>1$ such behavior emerges with the only difference that the
critical point is tuned into a new one.

\begin{figure}
\begin{center}
\includegraphics[width=8cm]{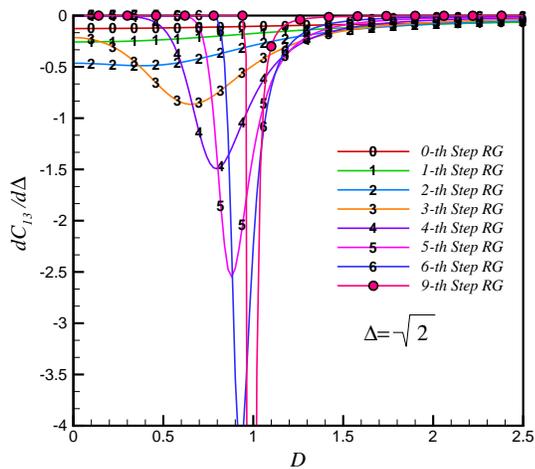}
\caption{(color online) First derivative of entanglement entropy and
its manifestation towards divergence as the number of RG iterations
increases (Fig.\ref{fig2}).} \label{fig3}
\end{center}
\end{figure}
\begin{figure}
\begin{center}
\includegraphics[width=8cm]{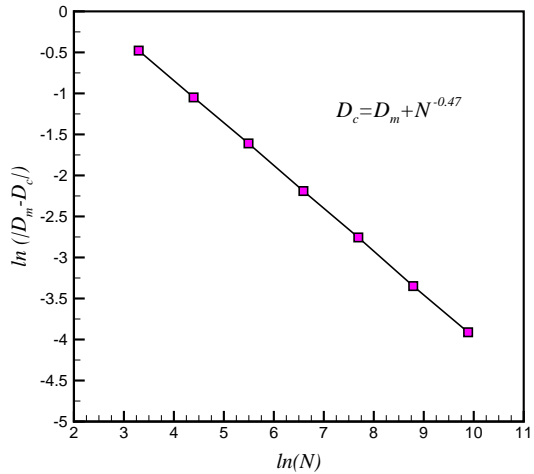}
\caption{(color online) The scaling behavior of $D_{m}$ in terms of
system size ($N$) where $D_{m}$ is the position of minimum in
Fig.\ref{fig3}.} \label{fig4}
\end{center}
\end{figure}
\begin{figure}
\begin{center}
\includegraphics[width=8cm]{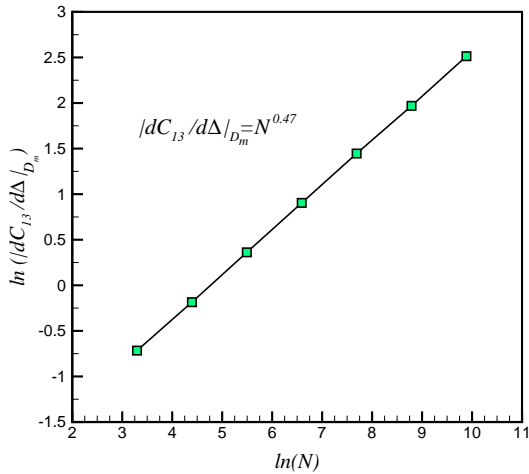}
\caption{(color online) The logarithm of the absolute value of
minimum, $\ln(\mid dC/d\Delta\mid_{min})$, versus the logarithm of
chain size, $\ln(N)$, which is linear and shows a scaling behavior.
Each point corresponds to the minimum value of a single plot of
Fig.\ref{fig3}.} \label{fig5}
\end{center}
\end{figure}

Further insight on the non-analytic behavior can be probed through
the diverging of the first derivative of the entanglement at the
critical point as long as the the thermodynamic limit is approached.
Plots related to the derivative of the entanglement at different RG
steps have been shown in the Fig.(\ref{fig3}). Each plot reveals a
minimum which becomes singular as the critical point is touched. At
the limit of large sizes of the model the singular behavior of the
entanglement becomes more pronounced. One may wonder how such
emerging singularity connects to the critical exponents or
universality class of the model. To this purpose, we shall see that
how position of minimum $D_{min}$ and minimum value itself $\mid
\frac{dC}{d\Delta}\mid_{D_{min}}$ scale with enlarging the size $N$
of the system. Such a computation determines the scaling law of
entanglement in one-dimensional spin systems and explicitly uncovers
an accurate correspondence with the critical properties of the
model. The position of the minimum ($D_m$) of $\frac{dC}{d\Delta}$
tends towards the critical point like $D_{m}=D_{c}-N^{-0.46}$ which
has been plotted in Fig.\ref{fig4}. Moreover, we have derived the
scaling behavior of $y\equiv|\frac{dC}{d\Delta}|_{D_m}$ versus $N$.
This has been plotted in Fig.\ref{fig5} which shows a linear
behavior of $ln(y)$ versus $ln(N)$. The exponent for this behavior
is $\mid\frac{dC}{d\Delta}|_{D_m} \sim N^{0.46}$. It should be
emphasized that this exponent is directly related to the correlation
length exponent, $\nu$, close to the critical point. It has been
shown in Ref.[\onlinecite{kargarian1}] that
$\mid\frac{dC}{d\Delta}\mid_{D_{c}}\sim N^{1/\nu}$  and
$D_{m}=D_{c}-N^{-1/\nu}$.

\begin{figure}
\begin{center}
\includegraphics[width=8cm]{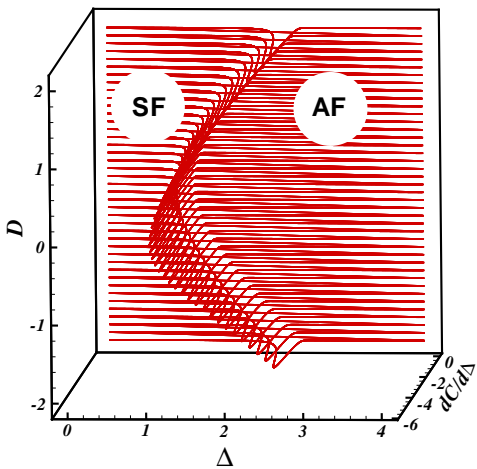}
\caption{The crack appearing in the derivative of the entanglement
corresponds to the critical line of the model that separates
antiferromagnetic (right) and spin fluid (left) phases.}
\label{fig6}
\end{center}
\end{figure}

Singular behavior of $\frac{dC}{d\Delta}$ corresponds to phase
transition for any value of DM interaction. It exhibits a singular
curve at the transition points. The latter can be characterized by
analyzing the derivative of entanglement for all values of DM
interaction. As an example, in Fig.(\ref{fig6}) the derivative of
entanglement in a three dimensional view has been shown. Noticeably,
the divergencies in the derivative are in perfect correspondence
with the parameter's value at which the phase transition occur. The
crack in the figure is just the critical line separating
antiferromagnetic from the spin fluid phases.

\begin{figure}
\begin{center}
\includegraphics[width=8cm]{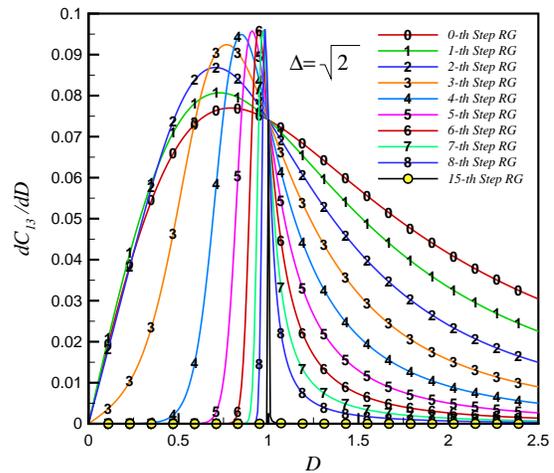}
\caption{The derivative of entanglement $\frac{dC}{dD}$ versus DM
interaction for fixed value of anisotropy $\Delta=\sqrt{2}$. Even in
limit of high steps of RG no singularity is observed.} \label{fig7}
\end{center}
\end{figure}

All above scaling functions hold for any value of anisotropy
parameter as long as $\Delta>1$, which is a direct result of the
fact that the parameter $D$ doesn't flow. This means that the
emerging DM interaction term in the model doesn't change the
universality class of the model. Thus far, taken the derivative of
entanglement with respect to anisotropy parameter, the singularity
appears at the critical point. To get more insight about the role of
DM interaction in the singularity of the entanglement, it is
convenient to plot the $\frac{dC}{dD}$ versus DM interaction as in
Fig.(\ref{fig7}). Even at the high steps of RG no singularity
detect. Observe that the pair $(D=1,\Delta=\sqrt{2})$ stands for a
point of singularity of derivative of entanglement with respect to
$\Delta$ as in Fig.(\ref{fig3}). However, there is no any signature
of the divergence in latter quantity at this point when the
derivative is taken with respect to $D$. This, again, verifies that
the DM interaction have not any thing to do with the universality
behavior of the model.

Indeed, for $\Delta>\sqrt{1+D^{2}}$ the long range behavior of the
model fall into the the universality class of the Ising model that
underlines the the appearance of the antiferromagnetic long range
order\cite{jafari2}. We emphasize that here, the critical line
$\Delta_{c}=\sqrt{1+D^{2}}$ from the Ising phase, i.e. $D_{m}
\longrightarrow D^{-}_{c}$. This directly comes from the fact that
the Ising phase is a gapped phase. Approaching the critical point,
the gap is closed as $E_{g}\sim |D-D_{c}|^{\nu z}$, where $z$ is the
dynamical exponent. Since in the limit of large sizes of the system
the critical point is touched as $D_{c}-D_{m}\sim N^{-1/\nu}$, we
are left with the result that the gap of the Ising phase in the
proximity of the critical point scales as $E_{g}\sim N^{-z}$.
Whenever $\Delta<\sqrt{1+D^{2}}$, the model is gapless. This can be
realized through a simple canonical transformation to the well known
XXZ model\cite{alcaraz,aristov} with the anisotropy
$\tilde{\Delta}=\frac{\Delta}{\sqrt{1+D^{2}}}$. This implies that
the model falls into a gapless spin fluid phase when
$\tilde{\Delta}<1$.

\begin{figure}
\begin{center}
\includegraphics[width=8cm]{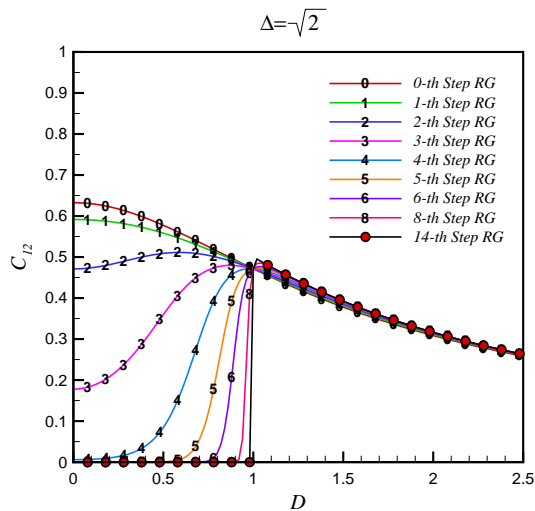}
\caption{The entanglement between first and second sites of the
block in terms of DM interaction at different RG steps. As before
the anisotropy parameter has been fixed at $\Delta=\sqrt{2}$.}
\label{fig8}
\end{center}
\end{figure}

Through this paper we have only considered the the entanglement
between two sites living on sides of a three-site block, i.e. the
middle site has bees traced out through the reduced density matrix.
We would like to emphasize that we could already consider the
entanglement between first two sites of the block. The entanglement
between first and second sites of the block has been shown that in
Fig.(\ref{fig8}). At zero-step of RG that represent a three-site
model, by increasing the DM interaction the entanglement between
sites reduce. This is in correspondence with behavior in
Fig.(\ref{fig2}) where the entanglement between first and third
sites is increased by increasing the DM interaction. This is
expected since when two parties get more entangled, they restricts
their entanglement with third party and vis versa that is a
reminiscence of the monogamy property\cite{Coffman} of entangled
objects. For $D<1$ that corresponds to the Ising phase, the
entanglement between two sites is shaved out. RG equations runs the
anisotropy to infinity dictating the spins to align. In contrary,
for $D>1$ that corresponds to the gapless phase, all plots behave
independent of RG steps. Observe that for $D<1$ neither first and
second sites $(C_{12})$ nor first and third ones $(C_{13})$ in the
large RG steps are entangled. This may not be surprising as in this
limit the model is characterized by a polarized state. The situation
is different for the gapless phase where the quantum fluctuations
dominate the system suppressing the alignment of spins.

If we were to take the derivative of plots, again the singularity
reveals itself at the critical point. Although $C_{13}$ and $C_{12}$
present different behavior, they share in exhibiting the critical
behavior of the model.

\section{Summary and conclusions \label{conclusion}}
Condensed matter systems have received impetus from the concepts
developed in quantum information theory. The central to this is the
entanglement being a unique measure of the quantum correlations. In
this stream we studied the entanglement in a one-dimensional
magnetic system in which many physical properties of realistic
complex materials can be understood through it. This model is the
well known XXZ model supplemented by a magnetic term arising from
the spin-orbit coupling. The phase diagram of the model is
determined by the anisotropy and  Dzyaloshinskii-Moriya interaction
(DM) parameters. In a simple model consisting of only three qubits,
the increasing of anisotropy parameter favors the alignment of spins
antiferromagnetically yielding a product ground state without
entanglement. However, tuning the DM interaction tends to build an
entangled state and restores the spoiled entanglement. This reviving
of entanglement can be understood via the fact that the DM
interaction contributes the strong planar quantum fluctuations that
pose the alignment ordering.

The thermodynamic limit of the model realized through the
renormalization group approach. The renormalization group not only
allows us to derive the critical points as well as phase diagram of
the model but via implementation of the RG steps we are able to keep
track the variation of the entanglement as the size of the system
becomes large. RG equations imply that the DM interaction tunes the
critical point, i.e there is critical line instead a single critical
point. However, the universality class of the model is unaffected in
the presence of the DM interaction, which can be clearly seen from
both RG equations and the scaling we obtained for the entanglement.
The role of the DM interaction can be well understood by analyzing
and comparing the derivative of entanglement with respect to DM
interaction and anisotropy parameter. In the former case, even in
the large RG steps, no singularity is observed. This can also be
justified by mapping the model into the well known XXZ model using a
canonical transformation. The derivative of the entanglement diverge
at the all points of the critical line. The line singularity
corresponds to the phase boundary separating the antiferromagnetic
from the spin fluid phases. The singularity accompanies by some
scaling functions with a emerging exponent that is related to the
correlation length exponent close to the phase transition. We also
verified that being gapped or gapless is relevant to the crossing
the phase transition. Via the enlarging the size of the system, the
singularity becomes more pronounced and touch the critical line from
the gapped phase(antiferromagnetic). This phenomenon results the gap
of the phase scales with the size of the system governed by the
dynamical exponent.

\begin{acknowledgments}
This work was supported in part by the Center of Excellence in
Complex Systems and Condensed Matter (www.cscm.ir).
\end{acknowledgments}

\end{document}